\documentclass[aps,twocolumn,prb,showpacs]{revtex4}
\usepackage{graphicx}
\begin{document}

%\preprint{}

\title{
Hole Doping Dependence of the Coherence Length in $La_{2-x}Sr_xCuO_4$ Thin
Films }

\author{H. H. Wen, H. P. Yang, S. L. Li }
\affiliation{ National Laboratory for Superconductivity, Institute
of Physics and Center for Condensed Matter Physics, Chinese
Academy of Sciences,PO Box 603, Beijing 100080  }

\author{X. H. Zeng, A. A. Soukiassian, W. D. Si and X. X. Xi}
\affiliation{ Department of Physics, The Pennsylvania State
University, University Park, Pennsylvania 16802}

%\date{\today}

\begin{abstract}
By measuring the field and temperature dependence of magnetization
on systematically doped $La_{2-x}Sr_xCuO_4$ thin films, the
critical current density $j_c(0)$ and the collective pinning energy
$U_p(0)$ are determined in single vortex creep regime. Together with the published data of
superfluid density, condensation energy and anisotropy, for the
first time we derive the doping dependence of the coherence length or vortex core size
in wide doping regime directly from the low temperature data. It
is found that the coherence length drops in the underdoped region
and increases in the overdoped side with the increase of hole
concentration. The result in underdoped region clearly deviates
from what expected by the pre-formed pairing model if one simply
associates the pseudogap with the upper-critical field.

\end{abstract}

\maketitle

%\twocolumngrid

It is well established that the superconductivity in the cuprates
is originated from the insulating anti-ferromagnetic parent
compound by doping holes into the system. The superconducting
transition temperature $T_c$ increases in the underdoped region
and decreases in the overdoped side, showing a parabolic shape of
$T_c$ vs. doped hole concentration $p$, i.e.,
$T_c/T_c^{max}=1-82.6(p-p_c)^2$ with $p_c\approx 0.15-0.16$,
$T_c^{max}$ being the maximum $T_c$ at the optimal doping point
$p=p_c$. At the underdoped side, an incomplete gap, namely the
pseudogap\cite{Pseudogap}, has been observed near the Fermi
surface at a temperature $T^*$, which is much higher than $T_c$.
It has been a core and longstanding issue to know the correlation
between the pseudogap and the superconductivity. Based on the
Uemura plot\cite{Uemura} and other experimental data, Emery and
Kievelson\cite{EmeryKievelson} put forward a picture to attribute
the superconducting transition in the underdoped region to a
Bose-Einstein condensation, while that in the overdoped side a
BCS-type. This model suggests that the pseudogap corresponds to
the energy scale of pre-formed pairing ( perhaps in the sense of
resonating-valence-bond singlet pairing\cite{RVB} ) and the
superconducting condensation occurs at $T_c$ when the long range
phase coherence is established. This picture has been cited to
explain the anomaly strong Nernst effect in normal
state\cite{XuZA,WangYY1}, the time-domain optical
conductivity\cite{Orenstein}, the Andreev reflection
experiment\cite{Deutscher}, etc. although the upper temperature
for these effects are still far below $T^*$.

In the conventional BCS scenario, the coherence length $\xi$, or
the upper critical field $H_{c2}=\Phi_0/2\pi\xi^2$, represent the
pairing strength. Meanwhile, $\xi$ measures the size of Cooper
pairs and vortex core, and near $H_{c2}$ the normal vortex core
with diameter of 2$\xi $ overlap with each other. In the BEC-BCS
picture mentioned above\cite{EmeryKievelson}, the pairing and the
condensation occur as two steps leading to a non-trivial issue of
defining the superconducting coherence length. However, the vortex
core size can be determined from experiment directly. In order to
get a deeper insight to the underlying mechanism of cuprate
superconductors, it is necessary to determine the doping
dependence of the vortex core size or the superconducting
coherence length $\xi$. This has turned out to be a very difficult
task due to the very high $H_{c2}(0)$ in cuprates. In early 90s,
by measuring the temperature dependence of diamagnetization or
excess conductivity under different magnetic fields, many
groups\cite{Hc2Early1,Hc2Early2} determined the slope of $H_{c2}$
near $T_c$ based on the mean-field fluctuation
theory\cite{Scaling}. Then according to the theory of
Werthamer-Helfand-Hohenberg ( WHH )\cite{WHH}, the value
$H_{c2}(0)$ is derived by $H_{c2}(0)=0.69|dH_{c2}/dT|_{T_c}$. It
was found that the coherence length obtained in this way decreases
with the increase of p in the underdoped region\cite{Hc2Early1}.
Because the true $H_{c2}(0)$ is very high, the method based on the
WHH theory is indirect, one naturally argues whether the
$H_{c2}(0)$ determined in this way is meaningful. In this Letter
we present a new approach to derive the zero-temperature coherence
length or vortex core size directly by analyzing the vortex
collective pinning energy on systematically doped
$La_{2-x}Sr_xCuO_4$ thin films.

Seven $La_{2-x}Sr_xCuO_4$ films made by pulsed laser ablation
technique, with {(00l)} orientation and nominal composition x =
0.06, 0.10, 0.125, 0.15, 0.175, 0.20 and 0.25 have been measured.
It is found that $p\approx $ 0.07 for the sample $ x=0.06$ due to
easy oxygen absorption, and $p\approx x$ for other samples. The
thickness of the films are typically 100nm and all samples have
rectangular shape with the same size. The films are smooth and
uniform without any traces of screw dislocations as shown by the
atomic force microscope. It is understood that the vortex pinning
in these films are from dense point-like disorders\cite{Griessen}.
Details about the fabrication were published
previously\cite{Film}. A vibrating sample magnetometer(VSM 8T,
Oxford 3001 )is used to measure the magnetization hysteresis loop(
MHL ) of the films with the magnetic field perpendicular to the
surface. In Fig.1 we show the $T_c$ vs. $p$ of these films, which
can be roughly described by $T_c/T_c^{max}=1-82.6(p-p_c)^2$ except
for a small plateau near $p=1/8$ as widely known. Fig.2 shows the
temperature dependence of the magnetic critical current density
$j_c$ at 0.03 T determined from the MHLs via $j_c = A\Delta M$,
where $\Delta M$ is the width of the MHL, and $A$ is a $T$ and $H$
independent pre-factor being the same for all films since they
have the same shape and sizes. We intentionally choose the low
field data for the present work because the single vortex creep
condition is satisfied here\cite{Blatter}. In this
semi-logarithmic plot, it is clear that $j_c$ drops slowly with
increasing temperature in the low temperature region showing a
linear-like behavior $Logj_c=Logj_c(0)- \alpha T$ with $\alpha $
clearly depending on the hole concentration $p$.

\begin{figure}
\includegraphics[width=8cm]{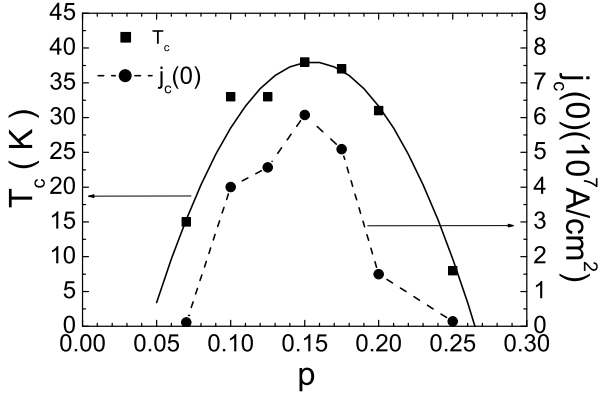}
\caption[] { Correlation between $T_c$ and $p$ ( squares ). The
solid line represents the empirical relation
$T_c/38=1-82.6(p-0.155)^2$. The circles stand for the $j_c(0)$
determined below. } \label{fig1}
\end{figure}

\begin{figure}
\includegraphics[width=8cm]{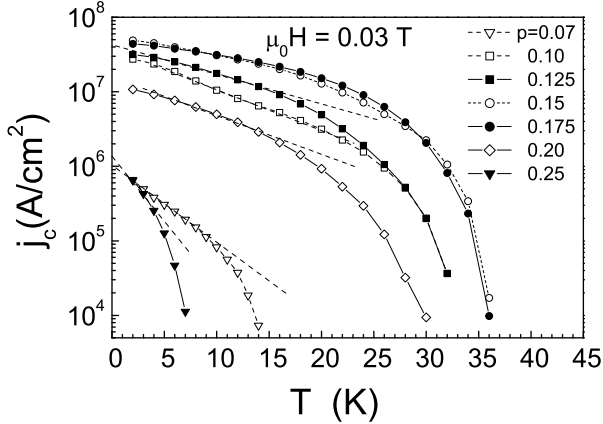}
\caption {Temperature dependence of the critical current density
$j_c$ determined from the MHLs at 0.03 T. The dashed linear lines
represent the the linear part for determining the slope $
dLogj_c/dT $ and the intercept $j_c(0)$ ( see text ).} \label{fig2}
\end{figure}

In addition to slight quantum tunnelling behavior below about 5 K,
the flux motion in HTS is dominated by the thermally activated
flux creep ( TAFC ) with the dissipative electric field
$E=v_0Bexp(-U/k_BT)$, where $v_0$ is the attempting speed for flux
creep, $U$ is the effective activation energy, $B$ the local
magnetic inductance. A general relation
$U(T,j_c(T))=(U_p(T)/\mu)[(j_c^0(T)/j_c(T))^\mu-1]$ is usually
assumed, where $\mu$ is the glassy exponent\cite{VG}, $U_p(T)$ is
the collective pinning energy, $j_c^0(T)$ and $j_c(T)$ are the
unrelaxed and the measured ( relaxed ) critical current density,
respectively. In the low temperature region, since $dlnj_c^0(T)/dT
\approx 0 $ and $U_p(T)\approx U_p(0)$, combining the TAFC
equation leads to

\begin{equation}
U_p(0)=-ln(v_0B/E)/[\frac{dlnj_c(T)}{dT}]_{T\rightarrow0}
\end{equation}

Here $ln(v_0B/E)$ depends on $p$ very weakly and can be taken as
constant ($\approx 10$)\cite{Wen95}. In this case $U_p(0)$ can be
determined by taking the slope of $Logj_c(T)$ vs. $T$ in low
temperature region.\cite{Quantum}. A linear fit to the low
temperature data gives rise to $U_p(0)$ ( the slope ) and $j_c(0)$
( the intercept ). The pinning energy $U_p(0)$ ( normalized at
$p=0.175$ ) is presented in Fig.3 and the zero-temperature
critical current density $j_c^0(0)=j_c(0)$ is shown in Fig.1. A
clear peak of $U_p(0)$ is observed at $p=0.175$ rather than at
$p=0.15$, as illustrated by a smaller slope of $dLogj_c(T)/dT$ for
the sample p=0.175 than for 0.15.

In the single vortex pinning regime, if the pinning is induced by
point-like disorders, like in our present thin films, the low
temperature collective pinning energy $U_p(0)$ is\cite{Blatter}

\begin{equation}
U_p(0)\propto \epsilon^{2/3} \xi^3H_c^2\sqrt{j_c(0)/j_0}
\end{equation}

where $\epsilon$ is the anisotropy ratio $\epsilon=m_{ab}/m_c$,
and the depairing current $j_0\propto H_c/\lambda\propto
\sqrt{E_c\rho_s}$ with $H_c$ being the thermodynamic field,
$\lambda$ the penetration depth, $E_c$ the condensation energy,
$\rho_s$ the superfluid density, and $E_c\propto H_c^2$. All
quantities in eq.(2) and the following discussions are valued at T
= 0 K, and for clarity we neglect the notes "0" for $E_c$ and
$\rho_s$, i.e., $E_c(0)=E_c$, $\rho_s=\rho_s(0)$. It is important
to note that eq.(2) is valid for both $\delta T_c$-pinning and
$\delta l$-pinning\cite{Blatter,Griessen}, pinnings induced by the
spatial fluctuation of $T_c$ and mean free
path respectively. Thus the coherence length $\xi$ can be obtained
via

\begin{equation}
\xi\propto \frac{U_p(0)^{1/3}\rho_s^{1/12}}{E_c^{1/4}j_c(0)^{1/6}\epsilon^{2/9}}
\end{equation}

In above equation, $j_c(0)$ and $U_p(0)$ are determined in our
experiment and shown in Fig.1 and Fig.3. The superfluid density
$\rho_s$ and the condensation energy $E_c$ for many systems have
been determined by Bernhard et al.\cite{Bernhard} and the
normalized values show roughly similar doping dependence in
different systems, thus we adopt these values in the following
analysis. For $p=0.07$ no data of $E_c$ is available, and we
follow the trend between $p=0.1$ and $p=0.14$, and extrapolate the
data down to zero at $p=0.05$ where $T_c$ vanishes ( shown by the
dotted line ). The doping dependence of the anisotropy ratio
$\epsilon$ is taken from the published values by Nakamura et
al.\cite{Nakamura} and Kao et al.\cite{Kao}. The computed results
of $\xi$ are shown in Fig.4 together with the experimental data
for YBCO-123 system\cite{Hc2Early1} determined by analyzing the
diamagnetization data based on the critical fluctuation
theory\cite{Scaling}. Here the calculated value of $\xi$ has been
normalized to $20\AA$ at $p_c=0.15$ according to the STM
measurement\cite{PanSH}. Clearly the coherence length $\xi$ drops
down with the increase of $p$ in underdoped region. It is
important to note that our approach is from the low temperature
data, but what turns out is very close to that from the critical
fluctuation analysis near $T_c$. The doping dependence of
$\epsilon$ does not seem to be very influential to the final
result as shown by the full circles with $\epsilon(p)$ considered,
and open circles with $\epsilon$ as a constant. Similar result is
obtained by Ando et al.\cite{Ando1} in analyzing the high field
magneto-conductivity in YBCO-123 system. In Fig.4 we also plot the
doping dependence of $\xi$ based on the preformed pairing model,
i.e., by assuming a simple linear relation
$H_{c2}(p)=H_{c2}(0)(1-p/0.28)$ and
$\xi=\sqrt{\Phi_0/2H_{c2}\pi}$, with $\mu_0 H_{c2}(0)=360T$. It is
clear that our result deviates from the expectation of the
preformed pairing model when simply relating the pseudogap as the
upper critical field. In the following we try to understand the
data within the BEC-BCS picture.

\begin{figure}
\includegraphics[width=8cm]{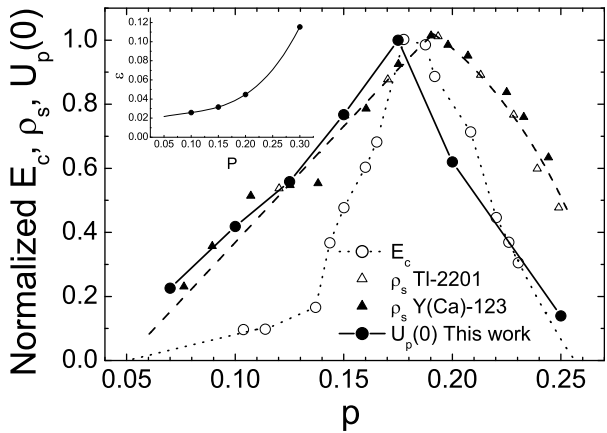}
\caption{Doping dependence of the normalized collective pinning
energy ( full circles ), the superfluid density $\rho_s$ (
up-triangles ) and the condensation energy $E_c$ ( open circles )
from ref.\cite{Bernhard}. The inset shows the doping dependence of
the anisotropy $\epsilon$ from
ref.\cite{Nakamura,Kao}}\label{fig3}
\end{figure}

\begin{figure}
\includegraphics[width=8cm]{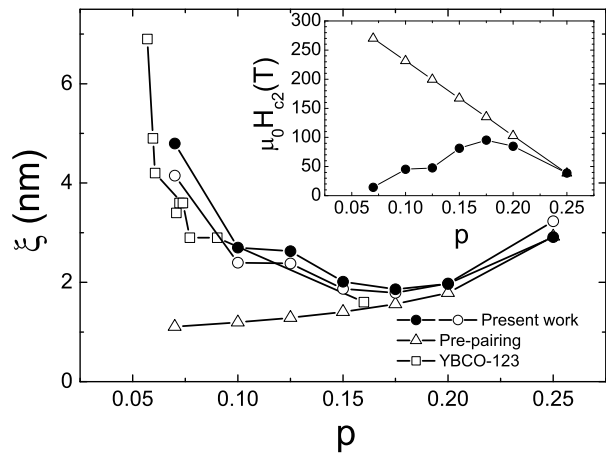}
\caption[]{ Computed coherence length ( full circles: without the
anisotropy ratio considered, open circles: with the anisotropy
ratio considered ). The up-triangles represent the case of
preformed pairing when simply associating  the upper critical
field with the pseudogap. The inset shows the upper critical field
derived in this work ( full circles ) and that expected by
equating the $H_{c2}$ with the pseudogap which is assumed to
decrease linearly with p.} \label{fig4}
\end{figure}

If the $H_{c2}(0)$ is defined as the critical field that destroys
the superconducting state at zero K, it should have very different
meanings within different schemes for the superconductivity. In
the conventional BCS picture, the energy scale for pairing, i.e.
the superconducting gap $\Delta_s$, is much smaller than the
energy corresponding to the phase stiffness $ E_{ph}\approx
\rho_s\hbar^2/m^*$, therefore the condensation occurs as soon as
the pairs are formed. The $H_{c2}$ in this case is understood as
the critical field for pair breaking. The condensation energy
$E_c=1/2N_F\Delta_s^2$, thus one can have a simple estimation that
$E_c\propto N_F\Delta_s^2\propto\rho_s\Delta_s$, where $N_F$ is
the electronic density of states near Fermi surface at $T_c$.
Since $\rho_s\propto 1/\lambda^2$, $H_{c2}\propto 1/\xi^2$,
$E_c\propto H_c^2\propto1/\xi^2\lambda^2$, one has then
$H_{c2}\propto \Delta_s$. Near $T_c$, the superconducting gap
vanishes, thus $H_{c2}$ becomes zero and the coherence length
diverges.

In the underdoped region, it is conjectured that the condensation
is BEC type with the Cooper pairs formed above $T_c$. If we assume
that $E_c=\rho_s\Delta_s$, since now the energy scale $\Delta_s$,
the pair breaking energy is much larger than $E_{ph}$, $H_{c2}$ is
now controlled by $\rho_s$. From the known
correlation\cite{Uemura} $\rho_s \propto T_c$, we can expect that
$H_{c2}\propto \rho_s \propto T_c$. This is very close to what we
obtained in the unerdoped region ( inset of Fig.4 ). It is
interesting to note that this simple relation $H_{c2}\propto T_c$
has also been obtained by Muthukumar and Weng\cite{Weng} within
the bosonic resonating-valence-bond theory. Following the BEC
picture for underdoped region, the vortex structure is also very
different from that expected from the BCS picture. Firstly, the
core state may be still gapped or partly gapped, reflecting the
pseudogap feature. This may explain why the zero bias conductance
peak ( ZBCP ) has not been observed within the vortex core by
STM\cite{PanSH,STMCore}, NMR\cite{NMRCore} and specific
heat\cite{WenCore}. Secondly, the pairing amplitude keeps almost
constant across the vortex core, while the superfluid density
$\rho_s$ now drops to zero in the core center and gradually
recovers its full value over a distance ( now the vortex core size
$\xi$). In this sense the vortex core size, or the newly defined
coherence length $\xi$ may be proportional to the London
penetration depth $\lambda$ although $\xi$ is much smaller than
$\lambda$. Thirdly, when $T_c$ is approached from below, both the
vortex core size or the coherence length and the penetration depth
will diverge due to the vanishing static superfluid density. At
low temperatures when the magnetic field is increased, the vortex
core tough each other at $H_{c2}$ in the conventional sense and
the extremely diluted superfluid density will not be strong enough
to sustain the long-life vortices. However the short-life vortices
can still be generated above $T_c(H)$ which gives rise to the
dynamic phase stiffness as observed by the ultra-fast optical
conductivity measurement\cite{Orenstein}. The anomalous strong
Nernst signal in the normal state can also be interpretated in
this way\cite{XuZA,WangYY1,WangYY2} although the upper magnetic
field for observing the Nernst signal may be more close to the
pair breaking field, which is proportional to the pseudogap. This
interpretation is consistent with the result of high field
measurement by Shibauchi et al.\cite{Shibauchi} who found that the
pseudogap can be closed by the Zeeman splitting under a high
magnetic field, while the field corresponding to the
superconducting peak on the c-axis magneto-resistance $H_{sc}$
scales with $T_c$ in the entire doping regime.  From our present
data and analysis, we cannot evaluate whether there is a quantum
critical transition of ground states\cite{Bernhard}, or the
transition concerning only the pairing symmetry\cite{Dagan} near
p=0.19. However in any case, in the overdoped side the pairing gap
drops with $p$ and the BCS picture is restored. Our results on the
doping dependence of the vortex core size call for future work on
observing the vortex structure directly by scanning tunnelling
microscope.

In conclusion, the hole doping dependence of the coherence length
or vortex core size has been determined directly from the low
temperature data in wide doping regime. It is found that the
coherence length drops in the underdoped region and increases in
the overdoped side with the increase of hole concentration. This
behavior clearly prevents the idea of relating the pseudogap with
the upper-critical field in underdoped region. The BEC-BCS picture
can be applied to interpret the data leading to consequences about
a new structure and many novel features of the vortex line.

\begin{acknowledgments}
This work is supported by the National Science Foundation of China
(NSFC ), the Ministry of Science and Technology of China, and the
Knowledge Innovation Project of Chinese Academy of Sciences. The
work at Penn State is supported in part by NSF under Grant No.
9623889 and ONR under Grant No. N00014-00-1-0294. We thank J. L.
Tallon for kind permission of using their published data ( Ref.20
). We are grateful for fruitful discussions with Z. Y. Weng, T.
Xiang and Q. H. Wang.
\end{acknowledgments}

%\bibliography{Coherencelength}

\begin{thebibliography}{}
\bibitem{Quantum}Due to the slight quantum tunneling effect below 5 K, we determine $U_p(0)$ by taking a linear fit to the data from 5 K to 15 K for $p$=0.10 to 0.20. The slight bending down of $Logj_c$ below 5 K is just due to the
quantum tunneling effect.  For the doping p=0.07 and 0.25, the
thermally activated flux creep is dominant even down to the lowest
temperature ( 2 K ) here due to very low $T_c$, therefore we use
the linear fit of low temperature data,
 as shown by the dashed lines. The slope gives rise to the $U_p(0)$ according to eq.1, and the intercept at T = 0 K
 gives directly $j_c(0)$.
\end{thebibliography}

\end{document}